\documentclass{aa}  

\usepackage{graphicx}
\usepackage[version=3]{mhchem}
\usepackage{tabularx}
\usepackage{xcolor}
\usepackage[normalem]{ulem}
\usepackage{txfonts}
\begin{document}

   \title{Investigating black hole accretion disks as potential polluter sources for the formation of enriched stars in globular clusters}

\author{Laurane Fréour\thanks{\email{laurane.freour@univie.ac.at}}\inst{1}
    \and Alice Zocchi\inst{1}
    \and Glenn van de Ven\inst{1}
    \and Elena Pancino\inst{2}
    }


\institute{Department of Astrophysics, University of Vienna, T\"urkenschanzstrasse 17, A-1180 Vienna, Austria 
\and INAF- Osservatorio Astrofisico di Arcetri, Largo Enrico Fermi 5, I-50125 Firenze, Italy
}

\date{Received ... / Accepted ...}
\titlerunning{Investigating BH accretion disks as potential polluter sources for the formation of enriched stars in GCs}
 
  \abstract
   {Accretion disks surrounding stellar mass black holes have been suggested as potential locations for the nucleosynthesis of light elements, which are our primary observational discriminant of multiple stellar populations within globular clusters. The population of enriched stars in globular clusters are enhanced in \ce{^{14}N}, \ce{^{23}Na}, and sometimes in \ce{^{27}Al} and/or in \ce{^{39}K}.
   In this study, our aim is to investigate the feasibility of initiating nucleosynthesis for these four elements in black hole accretion disks, considering various internal parameters such as the temperature of the gas and timescale of the accretion.
   To achieve this, we employed a 132-species reaction network. We used the slim disk model, suitable for the Super-Eddington mass accretion rate and for geometrically and optically thick disks. We explored the conditions related to the mass, mass accretion rate, viscosity, and radius of the black hole-accretion disk system that would allow for the creation of \ce{^{14}N}, \ce{^{23}Na}, \ce{^{27}Al}, and \ce{^{39}K} before the gas is accreted onto the central object. This happens when the nucleosynthesis timescale is shorter than the viscous timescale.
   Our findings reveal that there is no region in the parameter space where the formation of \ce{^{23}Na} can occur and only a very limited region where the formation of \ce{^{14}N}, \ce{^{27}Al}, and \ce{^{39}K} is plausible. Specifically, this occurs for black holes with masses lower than 10 solar masses ($m<10~M_\odot$), with a preference toward even lower mass values ($m<1~M_\odot$) and extremely low viscosity parameters ($\alpha <10^{-3}$). Such values are highly unlikely based on current observations of stellar mass black holes. However, such low mass black holes could actually exist in the early universe, as so-called primordial black holes.
   In conclusion, our study suggests that the nucleosynthesis within black hole accretion disks of elements of interest for the multiple stellar populations, namely, \ce{^{14}N}, \ce{^{23}Na}, \ce{^{27}Al}, and \ce{^{39}K} is improbable, but not impossible, using the slim disk model.
   Future gravitational wave missions will help constrain the existence of tiny and light black holes.}

   \keywords{Stars: abundances -- globular clusters:general -- globular clusters: individual: NGC~6752 -- Stars: black holes
               }

\maketitle


\section{Introduction} \label{sec:intro}
The origin of multiple stellar populations (MSPs) in globular clusters (GCs) is still an unresolved puzzle.
These populations appear as separate sequences in color-magnitude diagrams and they also show very specific patterns in the distribution of stellar abundances, the three most studied ones being the anti-correlation between carbon (C) and nitrogen (N), oxygen (O) and sodium (Na), and magnesium (Mg) and aluminium (Al).
Two main stellar populations have been identified: so-called ``pristine stars,'' with pristine chemical abundances, and ``enriched stars,'' showing enhancement in N, Na, and Al, and depletion in C, O, and Mg.

At first, \citet{1985Norris} found that giant stars with a high nitrogen abundance are also enhanced in sodium. \citet{1989Denisenkov,1990Denisenkov} mentioned evolutionary mixing to explain these chemical peculiarities. Sodium could be synthesized through the \ce{^{22}Ne}($p$,$\gamma$)\ce{^{23}Na} reaction in red giants, at the same location inside the star, where N is created and O destroyed, thus resulting in an overabundance of both N and Na and a depletion of O. Later on, \citet{1995Langer} suggested that Al could be considerably enhanced through very deep mixing and the activation of the MgAl reaction chain within bright giants. However, observations \citep{2004Gratton} have provided evidence against this type of evolutionary model as main sequence stars were found to also exhibit abundance variations \citep{1998Cannon}.
        
Many other theories have been suggested to shed light on the origin of these abundance patterns (see \citealt{2018Bastian} for an extensive review). Most of them involve a mixing between ``pristine'' and ``enriched'' material coming from different polluters. Already 40 years ago AGB stars were found to show abundance anomalies in some chemical elements  \citep{1981Cottrell}. Some 20 years later, \citet{2001Ventura} proposed that low-mass stars might have been polluted by the envelope of AGB stars, leading to the observed abundances; p-capture elements can indeed be synthesized during the hot-bottom burning phase in AGBs. However, \citet{2014Denissenkov} highlighted that the temperatures reached during this phase are too high ($T \gtrsim 100$ MK) to reproduce the observed isotopic ratios of Mg, requiring a narrow range of temperature around 80~MK, as shown by \citet{2007Prantzos}.
\citet{2014Denissenkov} suggested that supermassive stars \citep{2014Denissenkov_Hartwick} would be suitable polluter candidates as these stars can reach the needed central temperature, contrary to main sequence stars with a mass  of $M\lesssim 10^3 \rm M_{\sun}$ (such as fast-rotating massive stars, \citealt{2007Decressin}). Still, such supermassive stars currently remain hypothetical and have not been observed yet. \citet{2009DeMink} suggested interacting binaries as a polluter source, where the primary star could eject most of its envelope in the surrounding medium due to non-conservative mass transfer. Using interacting binaries and fast-rotating massive stars, \citet{2013Bastian} proposed the early disk accretion scenario. Low-mass protostars would sweep up the enriched material ejected by these two polluters through their protoplanetary disk. However, hydrodynamical simulations carried out by \citet{2016Wijnen} revealed an incompatibility between the different timescales involved. Protoplanetary disks would evaporate much faster than the timescale required for the creation of a sufficient amount of enriched material. None of the polluter sources proposed so far have been able to reproduce all the observational constraints on MSPs at the same time.

In particular, most of them are facing the so-called mass budget problem. Observations have revealed a high percentage of enriched stars, ranging from 40\% in low-mass clusters to 90\% in high-mass clusters  \citep{Milone2016}. The amount of enriched material required to form the enriched stars largely exceeds the amount of material that can be created by the polluters. To solve this issue, a modified IMF has been invoked \citep{2006Prantzos,2014Charbonnel} as well as a heavy loss of polluters from the GCs \citep{2013Renzini}, but both solutions are controversial \citep{2015Bastian}.

\cite{2018Breen} suggested that light elements could be created in accretion disks around black holes if the right conditions in terms of temperature and density of the gas were reached. These chemical elements could then be expelled through outflows from the accretion disk, mix with the surrounding pristine gas, and form the enriched stars in globular clusters. 
A sufficient amount of enriched material could be created in only $\sim 3$~Myr, depending on the initial number of black holes, assuming a super-Eddington accretion flow, overcoming the mass-budget problem.
The observed abundance trends in MSPs are consistent with H-burning through the activation of the CNO cycle as well as the NeNa and MgAl chains \citep{1993Langer}. The temperature involved in such reactions is a key parameter to obtain the expected yields. 

To reproduce the light element abundance patterns, the polluter should have a narrow range of temperature between 70~MK and 80~MK, \citep{2007Prantzos}. Heavier elements such as K require higher temperatures, namely, $T>180$~MK, to be created, which can often be incompatible with the proposed stellar polluter sources \citep{2016Iliadis, 2017Prantzos}. 
Black hole accretion disks can have more versatile properties than stars, depending on the optical depth, the viscosity, and the state of the black hole, resulting in broad ranges of densities, temperature, and radial velocities (see \citealp{2013Abramowicz} for a review of black hole accretion disks). 
A temperature profile covering a broad range could be an advantage for the creation of the enriched material. If the accretion disk has a radial temperature profile between 200~MK and 50~MK, it would be possible to synthesize not only Na, O, C, N, Mg, and Al, but also heavier elements like K. The diverse range of physical properties exhibited by accretion disks may account for the variety of properties among stellar populations in different globular clusters.

Several studies have set their primary focus on investigating heavy element nucleosynthesis within black hole accretion disks. \cite{1987Chakrabarti} explored the conditions under which nucleosynthesis could occur inside thick accretion disks around black holes. At the center of the thick disk, products of the PP-chain, the CNO cycle, and \textit{rp}-process (rapid proton process) could be created. They found that the amount of the elements produced was strongly dependent on the physical parameters of the disk, such as the mass of the black hole and the viscosity parameter, with smaller masses boosting the nucleosynthesis of such elements. This is promising as such reactions produce elements lighter than Z=40.
Using the formalism presented in \cite{1999Chakrabarti},
\cite{2000Mukhopadhyay} investigated three different types of accretion flow: hot, moderately hot and cooler flows. Depending on the inputs of their accretion disk model (seven variables in total), they find that the nucleosynthesis of some chemical elements is possible.
\cite{2008Hu} investigated the nucleosynthesis in advection-dominated accretion flow (ADAF) and outflow regions around a $10M_{\odot}$ black hole. They found that for a heating factor $f \approx 0.001$, representing the fraction of the viscously dissipated
energy, a significant amount of \ce{^{26}Al}, \ce{^{51}Cr}, \ce{^{53}Mn}, and \ce{^{55}Fe} could be created in the accretion disk.
These results suggest that nucleosynthesis could occur in accretion disks around black holes and be an effective way to increase the abundance of certain chemical elements in the interstellar medium. 

This work focuses on the issue of multiple stellar populations in globular clusters. As a starting point, our attention is only directed towards specific (light) elements whose abundance pattern is observed in most of the GCs: C, N, O, Na, Al, and Mg. We also considered the case of K, challenging current theories as it requires very high temperatures to be enhanced, which are often unreachable by the classical polluter sources.
Several points should be considered when reckoning whether black hole accretion disks could be the pollution source for the enriched material in globular clusters, for instance, whether the conditions in the temperature can be reached in an accretion disk; what the different timescales involved in the formation and accretion of the material are; and whether they are compatible with observations.
In this paper, we address the two first points. In Section~\ref{sec:nuc}, we introduce the reaction network that we employed and analyze the effects of temperature and time on the yields of C, N, O, Na, Al, Mg, and K in nucleosynthesis. Then, in Section~\ref{sec:accretion_disk}, we discuss the accretion disk model used and examine alternative models. Our approach for determining the feasibility of enriched material formation in an accretion disk surrounding a stellar mass black hole and the corresponding results are presented and discussed in Section~\ref{sec:method_results}. We summarize our findings in Section~\ref{sec:ccl}.


\section{Nucleosynthesis of the enriched material} \label{sec:nuc}
The stars in most GCs show chemical anomalies in light elements (He, C, N, O, Na, Al, and sometimes Mg). In other GCs, the MSPs also show a spread in heavier elements such as silicon (Si) or K.
In this study, we restricted the computation to C, N, O, Na, Al, Mg, Si, and K. The temperatures required to initiate the depletion of O and formation of N from the CNO cycle are $\ge 40$~MK, matching the temperature activating the NeNa chain. The MgAl chain requires higher temperatures $\ge 70$~MK to lead to an enhancement in Al and a depletion in Mg. Si and K are created for $T>100$~MK and $T>180$~MK, respectively.
Thus, we varied the temperature between 50~MK and 250~MK. 

The abundance of a given species over time is described by a differential equation that depends on its interactions with other chemical elements. 
If we consider two particles $j$ and $k$, then the mean lifetime of particle $j$ against destruction by particle $k$ is \citep{HIX2006}:
\begin{equation}
    \tau_k(j) = \frac{1}{\left\langle \sigma v \right\rangle_{j,k}n_k}
,\end{equation}
where $\left\langle \sigma v \right\rangle_{j,k}$ is the reaction rate between $j$ and $k$, depending on the temperature, and $n_k$ is the number density of particle $k$. Thus, a higher density leads to a lower mean lifetime for the particle.
In the reaction network simulations, this translates into unchanged final abundance when increasing (decreasing) the density by a certain factor and simultaneously decreasing (increasing) the simulation time by the same factor, as highlighted by \cite{2007Prantzos}.
Running two simulations with densities $\rho_1$ and $\rho_2 = 100 \times \rho_1$ for a time $t_1$ and $t_2=0.01 t_1$, respectively, leads to similar results.
Considering the coupling between density and time, we maintain a constant value for $\rho$ and investigate the evolution of the mass ratio of chemical elements with respect to time.
The main drivers in our simulations are thus the temperature and the timescale, $50<T<250~$MK and $10^{-6}<t<10^{14}~$s$~\approx 3~$Myr. The upper limit for the timescale corresponds to the approximate time needed for a cluster to be gas-free according to observations of young massive clusters \citep{2014Bastian}.

To run the nuclear reaction network simulations, we used the python package Pynucastro \citep{2022Clark}, an open-source project providing a user-friendly interface to solve nuclear reaction network equations.
For a two-body reaction involving species $A$ and $B$, the differential equation describing the evolution of the molar fraction, $Y_A$, of species $A$ is given by:
\begin{equation}
    \frac{d Y_A}{dt} = -(1 + \delta_{AB})\rho Y_A Y_B \frac{\mathcal{N_A} \left\langle \sigma v \right\rangle}{1 + \delta_{AB}} \ ,
\label{eq:Yab}
\end{equation}
where the term $(1+\delta_{AB})$ in the numerator accounts for double destruction in case species $A$ and $B$ are identical and $\mathcal{N_A}  \left\langle \sigma v \right\rangle$ is  the reaction rate. The reaction network is characterized by a set of differential equations similar to Eq.~\ref{eq:Yab}. Due to the stiffness of the system, specific numerical methods should be employed to solve such differential equations. We used the backward differentiation formula (BDF) integration method \citep{BDF} from the SciPy module.
Most reaction rates are taken from the REACLIB rate library \citep{Cyburt_2010}. In our reaction network, we included all the species producing or consuming the chemical elements involved in the CNO cycle, the MgAl, NeNa, and K reaction chains. In such way we are sure that no important reaction is omitted from the network. A total of 132 isotopes are included. 

In Sect.~\ref{subsec:4_enriched_mat}, we introduce the observed chemical abundances necessary to initialize our simulations and to compare our results with observations. 
As we are specifically interested in enhanced species (N, Na, Al, K), we define the enrichment timescale for the enhancement of these chemical elements in Sect.~\ref{subsec:4_timescales}. Such an enrichment timescale is particularly relevant when studying the formation of enriched material in accretion disks or in gaseous environments that are not stars, because the timescales associated with stellar objects are determined by stellar evolution models, and the nucleosynthesis in stars is different. Finally, we investigate  the conditions in temperature and timescale required to create the abundance patterns observed in GCs in Sect.~\ref{subsec:4_paramspace}. The work presented in this section is independent of the polluter source (AGBs, FRMS, black hole's accretion disks, ...) and of the formation scenario for multiple stellar populations.

\subsection{Observed pristine abundance pattern} \label{subsec:4_enriched_mat}

The first step to set up our simulations is to define an initial set of abundances (given as mass ratios). We used the pristine composition of NGC~6752 (see Table~\ref{tab:pristine_ab}) because this cluster has been widely studied over the past few years \citep{2013Carretta, 2016Lapenna, 2017Mucciarelli, 2017Pancino, 2018Lee, 2021Martins, 2019Mucciarelli}. It is one of the few globular clusters for which observations of C, N, Na, O, Al, Mg (including its isotopes), Si, and K are available \citep{2003Yong,2005Carretta,2007Carretta,2012Carretta}.
For these chemical elements, we estimated an average value for the pristine mass ratio.
For helium, we use a mass fraction of 0.246 \citep{2013Milone} and for iron we used $\left[ {\rm Fe/H}\right]=-1.5$, corresponding to $M_{\rm Fe}=4.7\times 10^{-5}$ \citep{2005Gratton}. For the other elements (Ar, Ca, Sc, and Ni), we proceeded as follows. First, we took the solar mass fraction of chemical element X from Table~4 in \cite{2021Lodders}. Then, as done by \cite{2007Prantzos}, following \cite{2000Goswami}, we recovered the abundance of element X at the metallicity of NGC 6752. These abundances are given as $\left[ {\rm X/Fe}\right]$ and we apply the following formula to convert from abundance to mass ratio:
\begin{equation}
\label{Eq:mass_ratio}
  M_{\rm X} = 10^{\left[\rm X/Fe\right]} \frac{M_{\rm Fe} M_{\rm X_{\sun}}}{M_{\rm Fe_{\sun}}}.
\end{equation}
To derive the abundances of isotopes for a chemical element, we further multiplied the right-hand side of Eq.~\ref{Eq:mass_ratio} by the isotopic ratio.
The adopted pristine mass ratios are presented in Table~\ref{tab:pristine_ab}.

\begin{table} 
        \centering
        \caption{ Pristine mass ratio used as input in the nuclear reaction code. The chemical elements are listed in the first column and their corresponding mass ratio in the second column.}
        \label{tab:pristine_ab}
        \begin{tabular}{cc}
                \hline
                \multicolumn{1}{c}{}
                Element \,  & $M_{\rm X}$\\
                \hline
                \ce{H} & 0.745  \\
                \ce{He}  &  0.246\\
                \ce{Li} & $1.6~10^{-9}$\\
                \ce{C} & $1.0~10^{-4}$  \\
                \ce{^{12}C} &$1.0~10^{-4}$ \\
                \ce{^{13}C}  & $1.0~10^{-6}$ \\
                \ce{^{14}N}  & $2.9~10^{-5}$  \\
                \ce{^{16}O}  & $1.0~10^{-3}$ \\
                \ce{^{20}Ne}& $1.0~10^{-4}$ \\
                \ce{^{22}Ne}& $1.6~10^{-5}$ \\
                \ce{^{23}Na} &  $1.2~10^{-6}$ \\
            \ce{^{24}Mg} &  $4.7~10^{-5}$  \\
            \ce{^{25}Mg} & $3.4~10^{-6}$   \\
            \ce{^{26}Mg} &  $3.4~10^{-6}$  \\
            \ce{^{27}Al} & $2.0~10^{-6}$ \\
            \ce{^{28}Si}&  $6.1~10^{-5}$  \\
                \ce{^{36}Ar}& $1.0~10^{-5}$    \\
                \ce{^{39}K}& $1.0~10^{-7}$  \\
                \ce{^{40}Ca}& $4.3~10^{-6}$  \\
                \ce{^{45}Sc}& $1.4~10^{-9}$  \\
                \ce{^{54}Fe}& $2.7~10^{-6}$ \\
                \ce{^{56}Fe}& $4.7~10^{-5}$  \\
                \ce{^{58}Fe} & $1.4~10^{-7}$ \\
                \ce{^{58}Ni}& $4.3~10^{-6}$ \\
                \hline
        \end{tabular}
\end{table}

\subsection{Enrichment timescale} \label{subsec:4_timescales}
\begin{figure}
\centering
\includegraphics[width=0.5\textwidth]{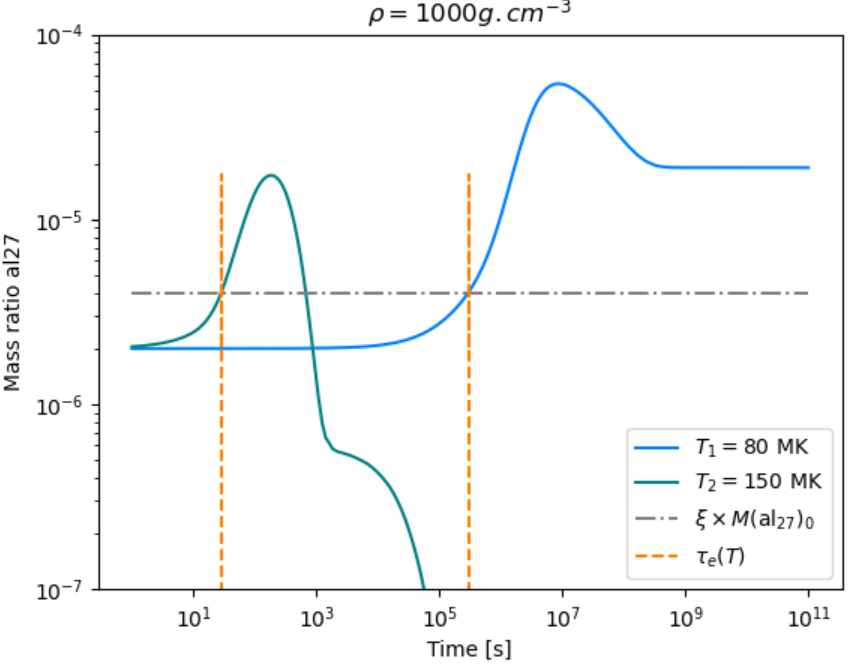}
\caption{Illustration of the concept of enrichment timescale. The green and blue lines represent the \ce{^{27}Al} mass ratio obtained from nuclear reaction simulations at temperatures of $T=80$~MK and $T=150$~MK, respectively. The dash-dotted line represents the threshold set to determine enrichment, with $\xi = 2$. In this case, \ce{^{27}Al} is considered enriched if $M_{^{27}Al} > 2 \times M_{{^{27}Al}_0}$. The two orange dashed lines correspond to the enrichment timescales, represented by their respective x-values.}
\label{fig:timescales}
\end{figure}

In nuclear physics, a widely used timescale is the burning time, $\tau^{\rm burn}$, characterizing the timescale on which an individual chemical element is destroyed \citep{HIX2006}:
\begin{equation}
    \tau^{\rm burn}=\frac{Y(Z)}{\dot{Y}(Z)}
    \label{eq:burn}
\end{equation}
where $Y(Z)$ is the nuclear abundance of specie $Z$ and $\dot{Y}(Z)$ the time variation of this abundance.
The burning time is proportional to the inverse of the density $\rho$ and the reaction rate $\left\langle \sigma v \right\rangle$:
\begin{equation}
    \tau^{\rm burn}\propto (\left\langle \sigma v \right\rangle \rho)^{-1}
.\end{equation}
However, for our purpose we are particularly interested in the timescale required for a chemical element to be enhanced with respect to the initial (or pristine) value.
Thus, we define the enrichment timescale, $\tau_e$.
We use the parameter $\xi$ to set a threshold for the definition of the timescale such that:

\begin{equation}
\begin{aligned}
\tau_{e} \equiv t(M_X> \xi ~M_{X_{0}}).
\end{aligned}
\end{equation}
The enrichment timescale of a chemical element, $X,$ is determined by such factors as $\xi$, the initial mass ratio, $M_{X_{0}}$, and the temperature of the simulation, the latest playing a crucial role in obtaining the mass ratio, $M_X$, in the nuclear reaction network. Figure~\ref{fig:timescales} illustrates the definition of $\tau_e$. We plot the evolution of the mass ratio of \ce{^{27}Al} for two temperatures. Then, we set the enrichment threshold $\xi =2$. Finally, we determine the enrichment timescale of \ce{^{27}Al} for each simulation by identifying the point in time when the \ce{^{27}Al} mass ratio surpasses twice the pristine mass ratio. If $\xi=2$, as in the case of Fig.~\ref{fig:timescales}, the enrichment timescale is thus the time necessary to obtain a mass ratio for the chemical element $X$ greater than two times the initial mass ratio, $M_{X_{0}}$.
Similarly, we define a depletion timescale in Eq.~\ref{eq:ts_depletion}, using the same parameter $\xi$. This timescale is used to inspect the conditions in temperature and time required to obtain the observed anti-correlations between chemical elements in MSPs: 
\begin{equation}
\begin{aligned}
\tau_{d}=t(M_X< \frac{1}{\xi} ~M_{X_{0}}).
    \label{eq:ts_depletion}
\end{aligned}
\end{equation}

\begin{figure*}[ht!]
\centering
\includegraphics[width=\textwidth]{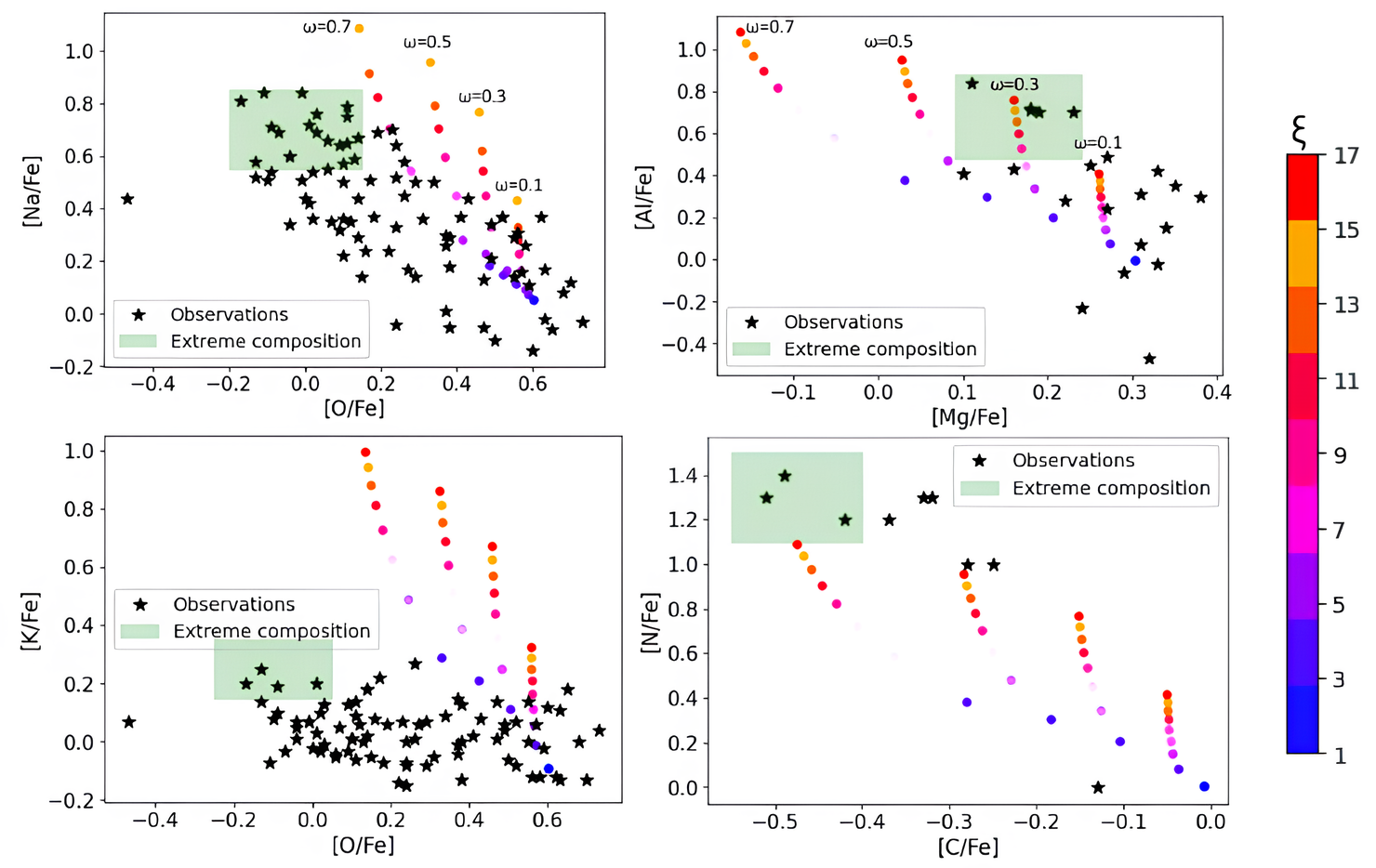}
\caption{Chemical abundance of O versus Na (upper-left), Mg versus Al (upper-right), O versus K (lower-left), and C versus N (lower-right). Black stars represent the observations taken from \cite{2005Carretta} for C and N, from \cite{2007Carretta} for O and Na, from \cite{2014Gruyters} for Al and Mg, and from \cite{2017Mucciarelli} for K. The green rectangle indicates the location of the most extreme abundances. The colored circles representing the abundances obtained using the reaction network ($T=80$~MK and $\rho = 1000$~g$\rm / cm^3$) and the dilution model introduced in Eq.~\ref{Eq:M2P} with $\omega = 0.1, 0.3, 0.5,$ and 0.7 are colored according to the enrichment level parameterized by the parameter $\xi$. While the extreme potassium abundances require only a small $\xi$, other chemical elements like oxygen require a large mixing weight and enrichment level to reproduce the extreme population.} 
\label{fig:epsilon_in_context}
\end{figure*}

\begin{figure}[ht!]
\centering
\includegraphics[width=0.45\textwidth]{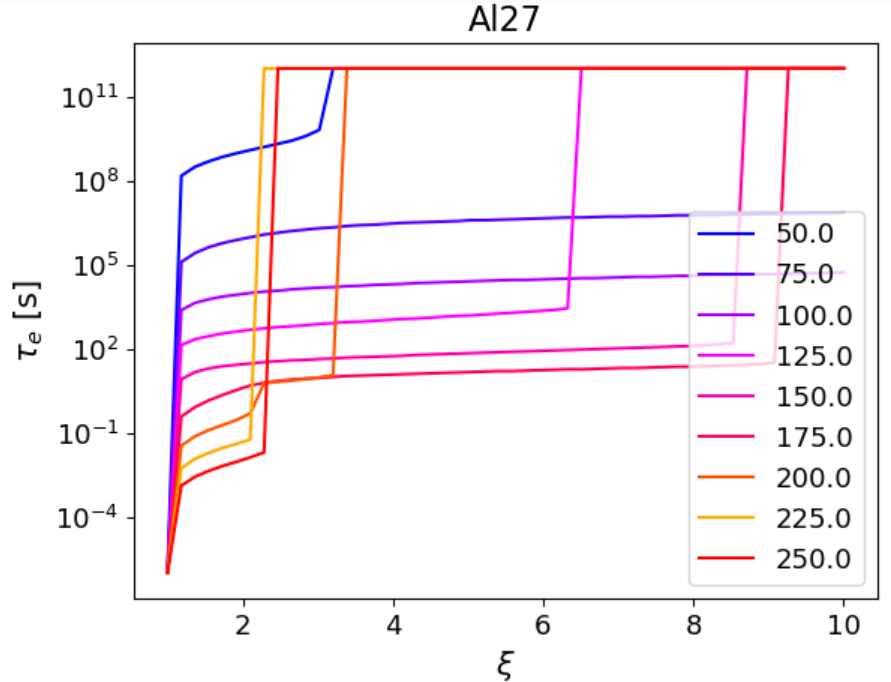}
\caption{Illustration of the evolution of the enrichment timescale, $\tau_e$, as a function of the parameter $\xi$. The lines are colored according to the temperature (in MK). The higher is $\xi$, the more enriched is the final composition in the corresponding chemical element, here \ce{^{27}Al}. The enrichment timescale is directly dependent on the temperature.}
\label{fig:xi_time}
\end{figure}

In order to provide some context for the values of $\xi$, we first need to introduce the notion of dilution. A widely accepted explanation for the emergence of MSPs in globular clusters involves the intervention of a polluter, which generates and expels enriched material containing a specific chemical element in a mass ratio denoted as $M_e$. This enriched material subsequently mixes with the surrounding pristine gas, characterized by a mass ratio of $M_p$, leading to the formation of an enriched star. The final composition of the resulting star is:
\begin{equation}
   M_{f} = (1-\omega) \, M_p + \omega \, M_e \ ,
    \label{Eq:M2P}
\end{equation}
where $\omega$ is the fraction of enriched material.
In Fig.~\ref{fig:epsilon_in_context} we plot the observed abundances of seven chemical elements taken from \cite{2005Carretta} for C and N, from \cite{2007Carretta} for O and Na, from \cite{2014Gruyters} for Al and Mg, and from \cite{2017Mucciarelli} for K, and represented as black stars. Our aim is to illustrate the influence of $\xi$ on the achieved enrichment after dilution to compare it with observations.
Here, we focus on the impact of $\xi$ and $\omega$ combined. The effect of $\omega$ alone has already been extensively discussed in previous works (e.g., \citealt{2007Prantzos,2011Di}). We used four distinct values for $\omega$, going from $\omega = 0.1,$ where the final star is mainly made of pristine material to $\omega = 0.7$, where the star is mainly made of enriched material. For each value of the mixing weight $\omega$ we vary the enrichment parameter $\xi$ from $\xi = 1$ (the enriched material has the same composition as the pristine) to $\xi = 17$ (enrichment of the material seventeen times higher than the pristine composition). The final abundances obtained are plotted as colored circles. We used the same value of $\xi$ for both chemical elements in the same panel. For example, in the upper left panel of Fig.~\ref{fig:epsilon_in_context}, the orange point on the dilution trail $\omega = 0.7$ corresponds to an enrichment threshold of $\xi = 13$ for Na and a depletion threshold of $\xi = 13$ for O.
An important observation from Fig.~\ref{fig:epsilon_in_context} is the discrepancy in abundance ranges among different chemical elements, with $\Delta [O/Fe]$ being approximately 1, while $\Delta [Mg/Fe]$ is around 0.3. Consequently, while a value of $\xi = 2$ (represented by blue dots, indicating that the expelled material has a mass ratio twice its pristine value) can reproduce the extreme magnesium and potassium compositions with a mixing weight of $\omega = $0.5, replicating the extreme oxygen abundance necessitates a significantly higher value of $\xi$ even with a substantial mixing weight. 

However, it is important to keep in mind that the output of our simulation, which is the mass ratio, is strongly influenced by the temperature. Consequently, the enrichment timescale is also affected by the temperature. Figure~\ref{fig:xi_time} illustrates how the enrichment timescale changes with the parameter $\xi$ for different temperatures, represented by the colored lines and ranging from $50$~MK (blue line) to $250$~MK (red line). When $\xi = 8$, the minimum value for the enrichment timescale is $\tau_e = 10$s, usually quite large compared to a typical viscous timescale (see Sect.~\ref{sec:accretion_disk}). Such value of $\xi$ only makes it possible to recover the extreme composition in Mg and Al, assuming a mixing weight $w>0.5$ (see Fig.~\ref{fig:epsilon_in_context}, pink dots). Overall, $\xi$ should be smaller than 3 to give an enrichment timescale $\tau_e < 1$s, and the temperature should also have specific values ($T>150$~MK). As shown on Fig.~\ref{fig:epsilon_in_context}, only the extreme Mg abundance can be recovered with $\xi = 3$ (blue dots).
The main conclusion drawn from the figure is that as the desired enrichment level increases, the corresponding timescale also increases for all temperatures. However, it is important to note that for certain temperatures (e.g., $T>200$~MK), achieving a high enrichment level (e.g., $\xi = 4$) is not possible within the duration of the simulation $\Delta t = 10^{14}$~s.

\subsection{Parameter space for the anti-correlations} \label{subsec:4_paramspace}
\begin{figure*}[ht!]
\centering
\includegraphics[width=\textwidth]{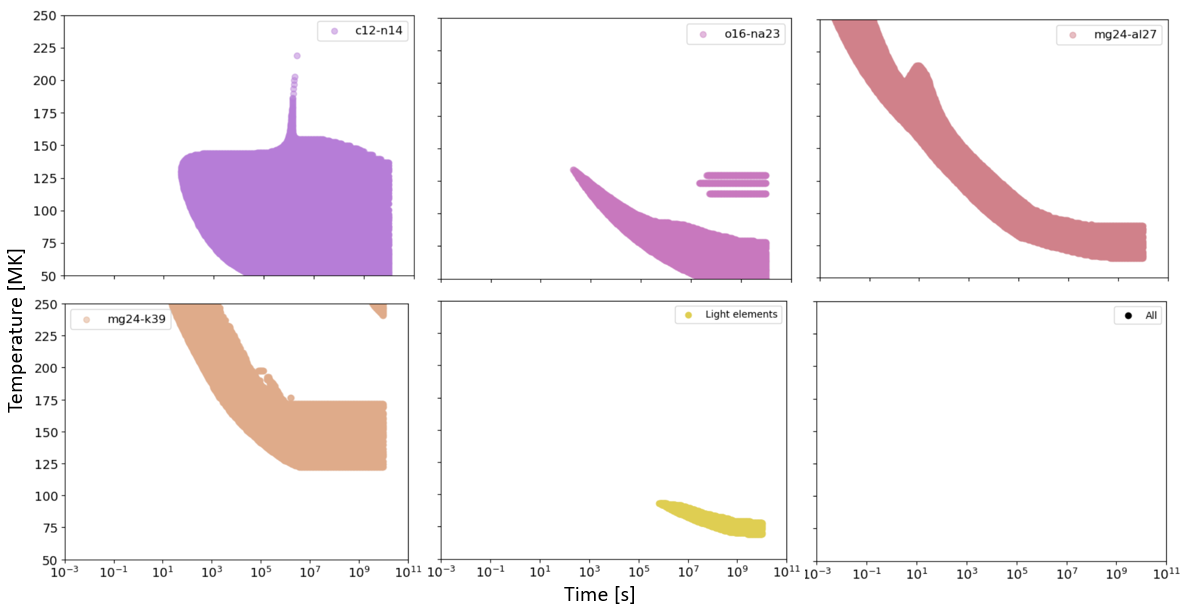}
\caption{Temperature and time generating the observed anti-correlations \ce{^{12}C}-\ce{^{14}N} (upper-left), \ce{^{16}O}-\ce{^{23}Na} (upper-right), \ce{^{24}Mg}-\ce{^{27}Al} (middle-left), and \ce{^{24}Mg}-\ce{^{39}K}  (middle-right). The lower middle and right panels represent the intersection of the parameter space reproducing the observed abundance patterns in light elements (middle) and in all the elements studied (right). Only a small region in temperature and time makes it possible to reproduce the light element patterns. At such temperature and time, \ce{^{39}K} can not be enhanced, so a second set of temperature and time values are needed.}
\label{fig:anti_cor}
\end{figure*}

Using the initial chemical abundances given in Table~\ref{tab:pristine_ab}, varying the temperature between $50~$MK$<T<250~$MK and the time $10^{-6}<t<10^{14}~$s, we obtain for each temperature and time the final mass ratios for the 132 chemical elements. The density is kept constant to $1000$~g $/ \rm cm^3 $. 
This value has been chosen arbitrarily and has only a small impact on the results, because of the coupling between time and density mentioned at the beginning of Sect.~\ref{sec:nuc}. If we would decrease the density, for instance to get a better agreement with accretion disks around compact objects in a binary system, the primary impact would be a rightward shift of the results shown in Figs.~\ref{fig:timescales} and \ref{fig:anti_cor}, toward larger timescales.
Using the definition of the enrichment and depletion timescales from Sect.~\ref{subsec:4_timescales} with $\xi=2$, we plot in Fig.~\ref{fig:anti_cor} the region of parameter space in temperature and time required to obtain the specified anti-correlations, key characteristic of MSPs. The empty parameter space in the lower right panel stands out. This means that in the entire temperature range explored and for $10^{14}$ order of magnitude in time, Potassium can never be produced together with the observed light chemical elements anti-correlations. For these light elements, only a very narrow range of temperature allows the simultaneous production of N, Na, Al, and Si and destruction of C, O, and Mg, in good agreement with \cite{2007Prantzos}. 
These results are valid regardless of the initial value for the density. As mentioned in Sect.~\ref{sec:nuc}, changing the density will only change the timescale in which chemical elements can be formed, thus shifting the parameter space above to the left or to the right.
These results, independent from polluter type and scenarios for the formation of MSPs, suggest a different origin for the heavy element patterns, pointing toward either multiple polluter types or a single polluter able to reach lower temperatures close to 70~MK as well as $T>150$~MK in a short amount of time. Motivated by these results we explore the concept of nucleosynthesis occurring in accretion disks surrounding stellar mass black holes as potential sources of pollution. We delve into this topic in the subsequent two sections.

\section{Accretion disk model} \label{sec:accretion_disk}
The Shakura-Sunyaev (S-S) disk model \citep{1973Shakura} is one of the most commonly used analytical model to describe accretion disk. Using a few assumptions, \citep{1973Shakura} showed that it is possible to solve analytically the system of thin disk equations. The S-S model is stable for sub-Eddington mass accretion rates ($\dot{m}<<\dot{m}_{Edd}$). The disk is geometrically thin (and is therefore referred to as a ``thin disk'' model), and optically thick.
The accretion disk is divided in three domains, depending on different physical conditions.
In the inner part, the radiation pressure and the electron scattering are dominant.
In the middle part, electron scattering is still the main source of the opacity, but gas pressure dominates.
Finally, in the outer part of the disk, the temperature and density are low. The gas pressure still dominates but the main contributor to the opacity is free-free absorption.

The Novikov-Thorne (N-T) model \citep{1973Novikov} is the relativistic generalization of the S-S model. The black hole is considered to be a rotating Kerr black hole. Relativistic correction terms are added to the Newtonian equations. These corrections are particularly important close to the inner edge of the rotating Kerr black hole where the gravitational pull is big.
Far away from the center, N-T equations are equivalent to the S-S equations.
In our case, we are mainly interested in the material close to the inner edge of the black hole,  where outflows arise. 

\citet{2018Breen} emphasized the necessity of a super-Eddington accretion flow to generate a sufficient amount of enriched material within a limited timeframe. This requirement stems from the need for the material to be produced in a matter of a few million years, aligning with the observational constraints observed in gas-free young massive clusters.
The slim disk model, originally introduced by \citet{1988Abramowicz} serves as a valuable expansion of the S-S model in scenarios involving a super-Eddington accretion flow, as visually represented in Fig.~1 from \citet{2013Abramowicz}. For accretion rates $\dot{m} \approx 0.1 \dot{m}_{Edd}$, the S-S disk may become unstable due to the radiation pressure contribution to the viscous torque \citep{1974Ligthman,1976Shakura,2002Janiuk}. For mass accretion rates close to or larger than the Eddington limit, the disk becomes thick and advection comes in the balance as an extra cooling mechanism. The slim disk can be used as an alternative disk model.

In an optically thick accretion flow and for high accretion rates (i.e., in the framework of the slim disk model), a portion of the photons produced within the disk becomes unable to exit its surface because they are confined by the inward flow of matter. This critical radius, at which this confinement happens, is commonly referred to as the ``trapping radius,'' $r_{trap}$. Such a radius is only dependent on the mass accretion rate. It delimits the accretion disk into two regions: the inner region (in the slim disk regime) and the outer region (where cooling is radiation-dominated),  and the accretion disk can be described by the standard model. The existence of such radius is one of the main difference with the advection-dominated accretion flow (ADAF) model in which the disk is optically thin and photons can therefore escape the disk immediately. As highlighted by \citet{1995Narayan}, who introduced self-similar solutions for ADAF, the temperature profile is then almost virial.
In such an accretion flow, the ion and electron temperature is decoupled at small radii.
The temperature profile plotted in their Fig.~3 matches the temperature criteria for nucleosynthesis at large radius $R>5000~R_{sch}$. 
We suggest that ADAF cannot be the process generating the enriched material in GCs for two main reasons. Firstly, the density is significantly low, on the order of magnitude much smaller than that of S-S disks, as illustrated, for instance, in \cite{2008Hu}. 
Given the interdependence of time and density in two-body reactions within a chemical reaction network (see Sect.~\ref{subsec:4_paramspace}), the generation of enriched material would necessitate an extensive duration to take place in an environment characterized by such a low density.
Secondly, the reaction chains involved in the formation of such material are activated for temperatures that are reached at very distant radius. As mentioned previously, outflows occur very close to the inner edge of the accretion disk. Thus, even if nucleosynthesis occurs in ADAF as suggested by \citet{2008Hu} and \citet{2017Zhang}, the synthesis of the chemical elements of interest for MSPs would happen at a large radius and could not be expelled into the surrounding medium to mix with the pristine gas. 

Therefore, we chose to use the slim disk model, covering accretion rates ranging from $10^{-4}~\dot{M}_{Edd}$ to $10^{5}~\dot{M}_{Edd}$. Although it has been acknowledged that the N-T model loses accuracy under super-Eddington accretion rates, we nonetheless included it in our simulations for verification purposes.
We describe the slim accretion disk by using the detailed expressions presented by \citet{2006Watarai}. Aspects that are of particular importance to our discussion include: the temperature, density, and radial velocity of the flow, $v_r$. 
The relevant equations used to describe the three quantities are characterized by two groups of equations, denoted by the subscript ${in}$ and ${out}$, depending on the position within the accretion disk, with respect to a threshold radius. Such radius threshold is set by the trapping radius, defined in \citet{2006Watarai} as the radius where the accretion time is shorter that the photon diffusion time, preventing photons from escaping due to the radial flow of matter. This radius marks the transition point between the standard disk, dominated by radiative cooling, and the slim disk, incorporating advection effects. The solutions within the trapping radius are as follows:
\begin{eqnarray}
    \left| v_{r_{in}} \right| &\approx& 3.17\times10^9 f \left(\frac{\alpha}{0.1}\right)\left(\frac{r}{r_g}\right)^{-1/2} ~~\rm cm~s^{-1},  \nonumber \\
    \Sigma_{in} &\approx& 2.36\times10^3f^{-1} \left(\frac{\alpha}{0.1}\right)^{-1} \left(\frac{\dot{m}}{100}\right)\left(\frac{r}{r_g}\right)^{-1/2}\rm~~g~cm^{-2}, \\
    T_{in} &\approx& 2.52\times10^7f^{1/8}\left(\frac{m}{10}\right)^{-1/4}\left(\frac{r}{r_g}\right)^{-1/2} ~~ \rm K \nonumber \ ,
\label{eq:eq_in}
\end{eqnarray}
where $\alpha$ is the viscosity of the gas, $r_g = 2GM/c^2$ is the Schwarzschild radius, and $m = M/M_{\odot}$ and $\dot{m} = \dot{M}/(L_E/c^2)$ are the normalized mass and mass accretion rate, expressed in unit of solar mass and Eddington accretion rate, respectively.
These equations are slightly different from the canonical self-similar solutions (see, e.g., \citealt{1999Wang}), because of the treatment of energy transport. In \cite{2006Watarai}, the ratio of the advective cooling rate to the viscous heating rate, $f,$ is not constant and given by:
\begin{equation}
f(\dot{m},r) = 0.5\left(D^2 x^2 + 2 - Dx\sqrt{D^2x^2 + 4}\right) \ ,
\label{eq:f}
\end{equation}
where $D$ is a numerical coefficient taken to be equal 2.18, and $x = $($r$/$r_g$)/$\dot m$.
The quantities outside the trapping radius are similar to the solutions from \cite{1973Shakura}:
\begin{eqnarray}
    \left| v_{r_{out}} \right| &\approx& 5.96\times10^{12}\left(\frac{\alpha}{0.1}\right)\left(\frac{\dot{m}}{100}\right)^2\left(\frac{r}{r_g}\right)^{-5/2} ~~\rm cm~s^{-1}, \nonumber \\
    \Sigma_{out} &\approx& 1.26 \left(\frac{\alpha}{0.1}\right)^{-1} \left(\frac{\dot{m}}{100}\right)^{-1}\left(\frac{r}{r_g}\right)^{3/2}\rm~~g~cm^{-2}, \\
    T_{out} &\approx& 4.79\times10^7\left(\frac{m}{10}\right)^{-1/4}\left(\frac{\dot{m}}{100}\right)^{1/4}\left(\frac{r}{r_g}\right)^{-3/4} ~~ \rm K. \nonumber \ 
\label{eq:eq_out}
\end{eqnarray}

It is important to highlight that the density is provided as (projected) surface density. The conventional conversion formula from 2D surface density to a 3D density profile is as follows:
\begin{equation}
    \Sigma = \int_{-H}^{H} \rho dz
.\end{equation}
Here, $H$ represents the disk's height. However, for the sake of simplicity and due to the coupling between time and density in nuclear reaction network simulations, we opted for the simplified vertical one-zone approximation commonly used in the standard disk model: $\Sigma = 2H \rho$. This choice is further discussed in Sect.~\ref{sec:ccl}. Finally, the viscous timescale can be defined as the time needed for the material at a given radius, $R,$ to be accreted onto the compact object:
\begin{equation}
    \tau^{\rm visc}=\frac{R}{v_r}
    \label{eq:visc}
.\end{equation}

\section{Nucleosynthesis in black hole accretion disks} \label{sec:method_results} 
In Sect.~\ref{sec:nuc}, we showed the challenge to simultaneously obtain the anti-correlations between C and N, O and Na, Mg and Al, and Mg and K from the same temperature range.
In this section, we aim to investigate whether the thermodynamic conditions of accretion disks are favorable for the production of N, Na, Al, and K.

\subsection{Method} \label{subsec:5_method}
Table~\ref{tab:method} summarizes the method followed in this work to obtain the quantity of interest represented by the viscous and enrichment timescales from the inputs of the accretion disk model.
Firstly, we coded the accretion disks' equations.
The outputs of the disk model are the density, temperature, and radial velocity of the accretion flow. The temperature is used as an input for the reaction network, as all the reaction rates are temperature dependent. By varying the parameters of the black hole and its accretion disk (black hole mass, $m$, mass accretion rate, $\dot{m}$, radial distance to the black hole, $R$, viscosity of the gas, $\alpha$), we obtained different densities, temperatures, and radial velocities for the gas. Using these temperatures and densities as input of the reaction network, we can compute the enrichment timescale and compare it to the viscous timescale in order to determine whether accretion disks can host the required nucleosynthesis of different (light) chemical elements.

\begin{table*}[]
    \centering
    \begin{tabular}{|p{0.9cm}|p{2.5cm}|p{2.7cm}||p{0.9cm}|p{2.5cm}|p{2.7cm}||p{2.5cm}||}
     \hline
     \multicolumn{3}{|c||}{Accretion disk model} & \multicolumn{3}{|c||}{Reaction network} &\multicolumn{1}{|c||}{Quantity of interest}\\
     \hline
     Input & Range explored & Output & Input& Range explored & Output & \\
     \hline
     $m$&[$10^{-4},10^8]~M_{\odot}$ &$T=f(m,\dot{m},R,\alpha)$& T & $ >2\times 10^6$~K &$M_X=f(M_{X_{0}},T,t)$& $\tau_{visc}=\frac{R}{v_R}$\\
     $\dot{m}$&[$10^{-4},10^5]~ \dot{M}_{Edd}$ &   $\rho=f(m,\dot{m},R,\alpha)$ &t &$[10^{-6},10^{14}]~$s &  & $\tau_{e}=f(M_X)$ \\
     $R$ &[$3,300]~R_{sch}$ & $v_R=f(m,\dot{m},R,\alpha)$ & & &  & \\
     $\alpha$ &[$10^{-4},1$]&  & & & & \\
    \hline
    \end{tabular}
    \caption{This table provides an overview of the methodology employed to obtain the quantities of interest presented in the last column. The first three columns provide the input parameters, the ranges explored, and the outputs of the accretion disk model. The three next columns present the inputs, their range, and the outputs of the reaction network used. Two steps lead to the computation of the important timescales: the viscous timescale only depends on the accretion disk model, while the burning/enriched timescale depends on the evolution of the mass ratio of chemical element $X$ (dependent on the temperature, thus, on the disk model and on the reaction network).}
    \label{tab:method}
\end{table*}

The range of mass, mass accretion rate, radius, and viscosity investigated is given in the second column of Table~\ref{tab:method}. The range of black hole masses and accretion rates is very broad, covering  11 and 8 orders of magnitude, respectively. To ensure the completeness of our investigation, we sampled our parameter space using a two-step method designed to optimally cover it. First, we divided the space into a regular grid of range of [$l \times 10^i$, $l \times 10^{i+1}$] for $i$ varying between 0 and $\log_{10}(u/l)-1$ , where $l$ and $u$ are the lower and upper bound of the interval.
Then, within each grid cell, we use latin hypercube sampling \citep{1979McKay} to take 100 points randomly sampled. A Latin hypercube is a sampling method that ensures each variable in a set is sampled uniformly across its range, and each sample is independent of the others.
We have a total of $11 \times 8 \times 2 \times 4$  different intervals within our four-dimensional (4D) parameter space. 
Within each interval, we randomly sampled 100 values for each parameter. We thus obtained a total of $704 \times 100 \times 4 = 281,000$ values of ($m$,$\dot{m}$,$r$,$\alpha$). 
We used these values as inputs for the analytical model of the accretion disk and obtained 281,000 tuples of ($T$,$\rho$,$v_r$). For the nuclear reaction network, we only used the tuple with $T>2$~MK as no nucleosynthesis of N, Na, Al, and K can occur for smaller temperatures. We are left with a total of 25,730 values. For each value we run the nuclear reaction code, solving a stiff system of equations (Eq.~3 in \citealp{2022Clark}). The results are presented in Sect.~\ref{subsec:5_results}. 

\subsection{Results} \label{subsec:5_results}
\begin{figure}[ht!]
\centering
\includegraphics[width=0.5\textwidth]{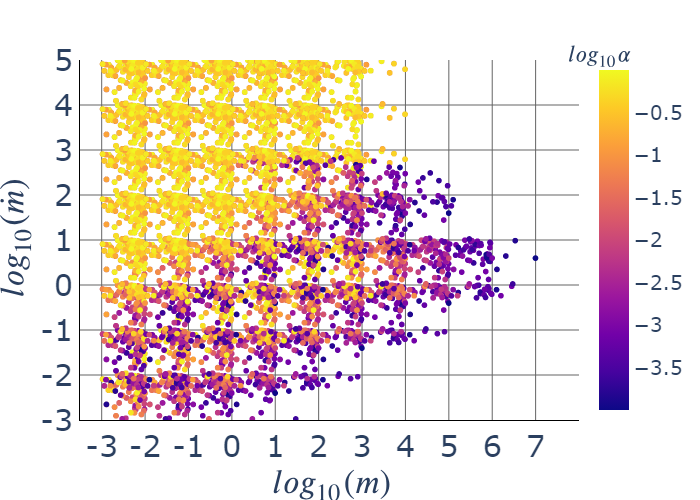}
\caption{Parameter space exploration for the black hole and accretion disk system. A total of 25,730 points are plotted out of the initial 281,000 parameter, representing approximately 9\% of the samples ($m$, $\dot{m}$, $r$, $\alpha$) that yield temperatures exceeding the threshold of $2$~MK required as input into the nuclear reaction network. The points are displayed in a logarithmic scale on the $m$ versus $\dot{m}$ plane and color-scaled based on $\alpha$ values.}
\label{fig:param_space_full}
\end{figure}

\begin{figure*}[ht!]
    \centering
    \includegraphics[width=0.49\textwidth]{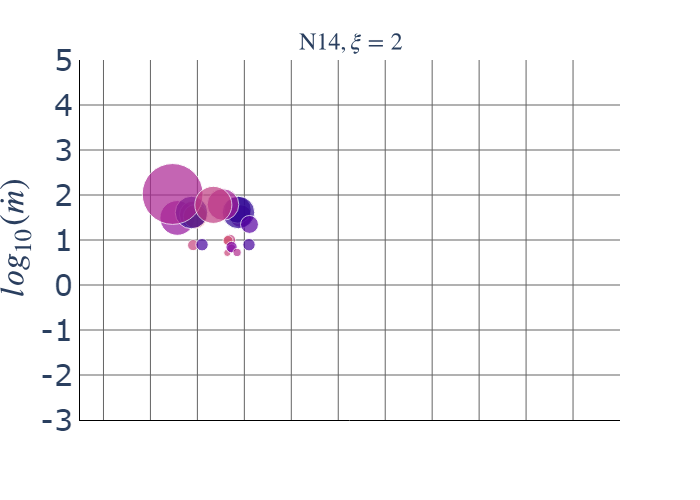}\hspace{-13mm}
    \includegraphics[width=0.49\textwidth]{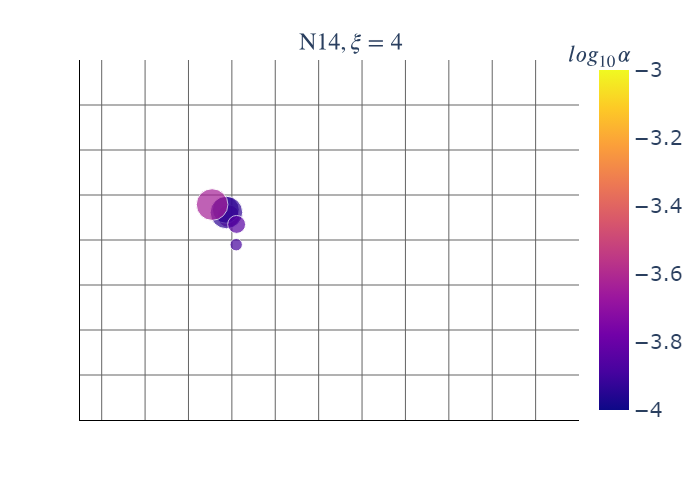}
    
    \vspace{-7mm} 
    
    \includegraphics[width=0.49\textwidth]{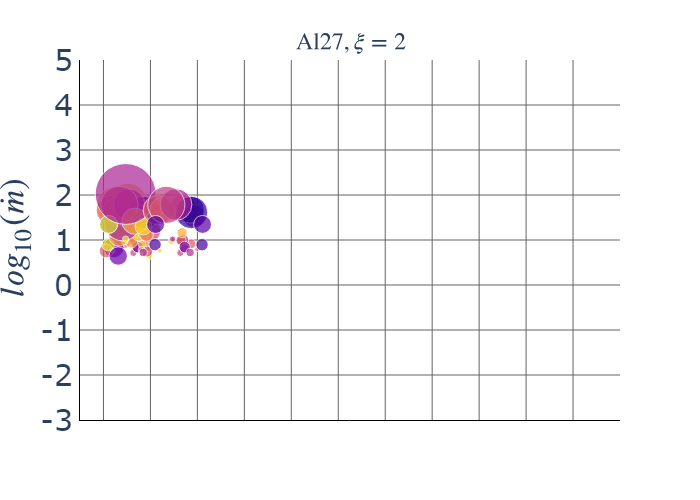}\hspace{-13mm}
    \includegraphics[width=0.49\textwidth]{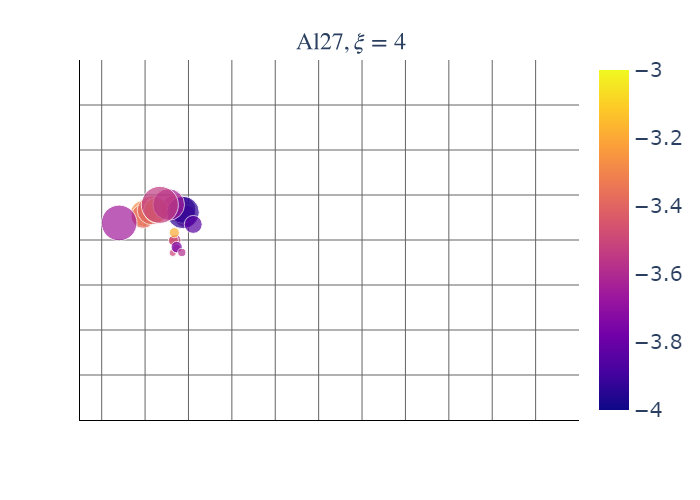}
    
    \vspace{-7mm} 
    
    \includegraphics[width=0.49\textwidth]{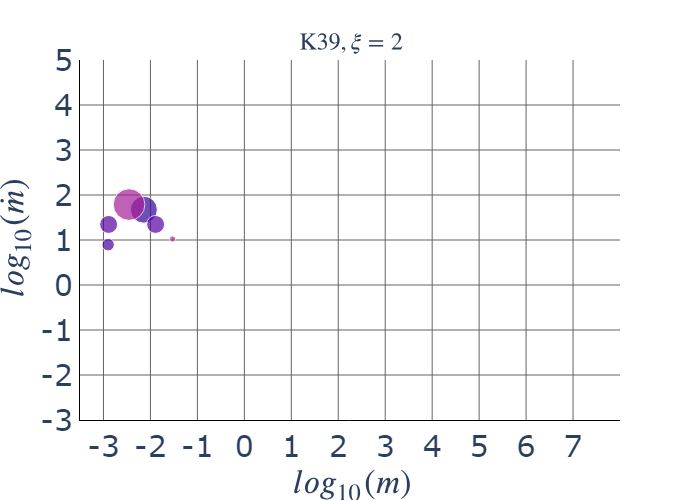}\hspace{-13mm}
    \includegraphics[width=0.49\textwidth]{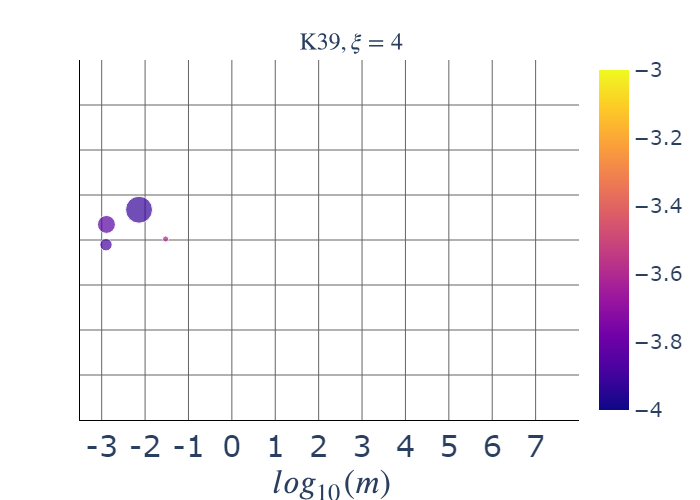}
    
    \caption{Identification of ($m$, $\dot{m}$, $r$, $\alpha$) values resulting in an enrichment of \ce{^{14}N}, \ce{^{27}Al}, and \ce{^{39}K} larger than two ($\xi = 2$, left panel), and greater than four ($\xi = 4$, right panel). The size of the points corresponds to the distance to the center of the black hole, $r$.
    Only 20 (upper-left), 82 (middle-left), and 6 (lower-left) out of the 25,730 points enable an enrichment by a factor of 2, translating into 0.07\%, 0.3\%, and 0.02\% cases of enrichment. When $\xi = 4$, the number of points becomes 19 (upper-left), 6 (middle-left), and 4 (lower-left), translating into 0.07\%, 0.02\%, and 0.015\% case of enrichment. The largest distance to the black hole making an enrichment possible occurs for a value of $r = 61 r_{\text{sch}}$, visible in the upper and middle left panels.}
    \label{fig:param_space_al}
\end{figure*}

In the following, we study the ratio $\tau_{nuc}=\frac{\tau_{\rm visc}}{\tau_{\rm e}}$. If $\tau_{\rm visc}<\tau_{\rm e}$, thus $\tau_{nuc}<1$, the material in the accretion disk is accreted before nucleosynthesis can occur. If $\tau_{nuc}>1$, the enrichment timescale is larger than the viscous timescale, and nucleosynthesis could occur in the accretion disk before the gas is accreted. 
The results are plotted in Figs. \ref{fig:param_space_full} and~\ref{fig:param_space_al}. 

Figure~\ref{fig:param_space_full} shows the entire range of parameter values leading to a temperature high enough for nucleosynthesis to occur ($T>2$~MK), in the logarithmic plane, $m$ versus $\dot{m}$, and color-scaled according to $\alpha$. Some combinations of parameters, such as a high black hole mass and low accretion rate, result in temperatures that are too low to allow for nucleosynthesis. We ran the simulation for different enrichment thresholds (values of $\xi$).
In the left panel of Fig.~\ref{fig:param_space_al}, we show the values of ($m$, $\dot{m}$, $r$, $\alpha$) resulting in an enrichment of \ce{^{14}N}, \ce{^{27}Al}, and \ce{^{39}K} greater than two times the pristine value ($\xi = 2$). Only a very light black holes with $m<1 M_{\odot}$ and very small viscosity parameter $\alpha < 0.003$ makes the enrichment of \ce{^{27}Al} and \ce{^{39}K} possible.
Regarding \ce{^{14}N}, a few models around $m \approx 1~M_{\odot}$ and moderate mass accretion rate $1~\dot{M_{Edd}}<\dot{m}<100~\dot{M}_{Edd}$ makes the enrichment possible. 
No combination of ($m$, $\dot{m}$, $r$, $\alpha$) create the right conditions in temperature and density to enrich \ce{^{23}Na} with a threshold of $\xi =4$. 

\begin{figure*}[ht!]
\centering
\includegraphics[width=0.48\textwidth]{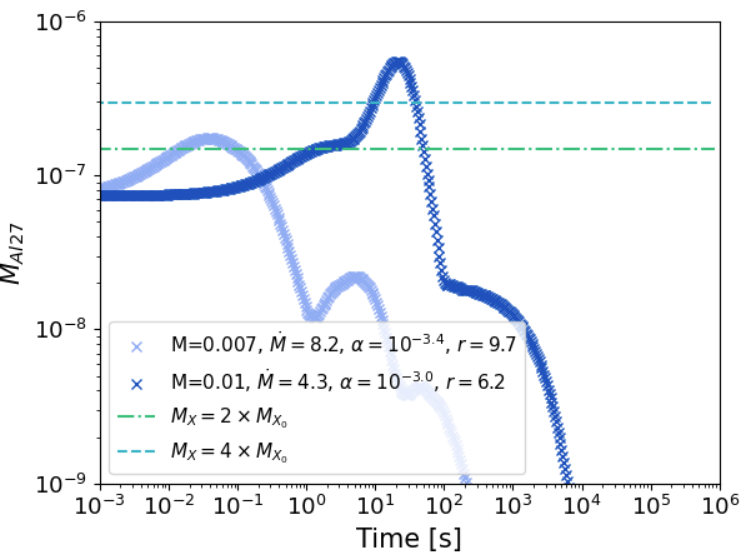}
\includegraphics[width=0.48\textwidth]{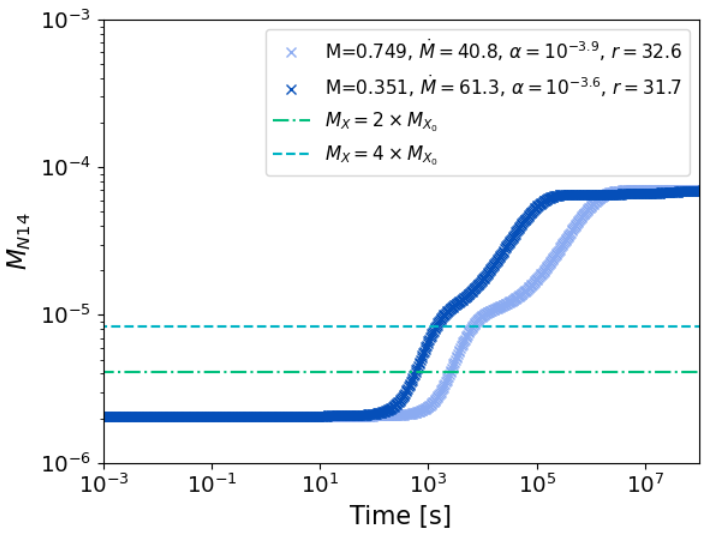}
\caption{Variation of \ce{^{27}Al} and \ce{^{14}N} mass ratio over time for two runs where the enrichment timescale exceeds the viscous timescale. The label details the parameters of the black hole and accretion disk system. The mass, $M,$ is expressed in $\rm M_\odot$, the mass accretion rate in terms of the Eddington rate, $\dot{M_{Edd}}$, and the radius in unit of the Schwarzschild radius. The two thresholds for considering different enrichment level are represented by the dash-dot and the dotted lines. While the \ce{^{27}Al} mass ratio only remain above the lines for a brief duration, the enrichment in \ce{^{14}N} is rather stable and remains at a high level until the end of the simulation.}
\label{fig:duration_enrichment}
\end{figure*}

We further investigated the potential formation of \ce{^{14}N} and \ce{^{27}Al} focusing on an important aspect that has not been addressed so far: the duration of the enrichment. In Fig.~\ref{fig:timescales}, the green-blue curve demonstrates a defined duration for the enrichment, which we estimate to be approximately $\Delta \tau_e \approx 100$ seconds. Conversely, the blue curve indicates an indefinite duration for the enrichment. If $\Delta \tau_e$ is small, the chemical element is very quickly destroyed. Thus, even if it would be created, it would be very unlikely that it can be expelled from the accretion disk, pollute the surrounding medium, and explain the abundances in enriched stars in globular clusters. 
We define the duration of the enrichment:
\begin{equation}
    \Delta \tau_{e} = \min_{t>\tau_e} ~t(M_X < \xi M_{X_0}) - \tau_e.
\end{equation}
Here, $\Delta \tau_e$ can be visualized in Fig.~\ref{fig:duration_enrichment}, showing the evolution of \ce{^{14}N} and \ce{^{27}Al} mass ratios as a function of time for two different successful runs, as the time length when the different colored curves are above the enrichment thresholds represented by the green and red lines. 
We compute the duration of the enrichment $\Delta \tau_e$ for each of the points shown in Fig.~\ref{fig:param_space_al}.
For \ce{^{27}Al}, $\Delta \tau_e$ barely exceeds 100 seconds and never exceeds 500 seconds in the 113 successful runs.
Regarding \ce{^{14}N}, once the mass ratio is exceeding the threshold, it stays at a constant value until the end of the simulation.

\subsection{Discussion} \label{subsec:diskussion}
The results presented in Section~\ref{subsec:5_results} indicate the potential for nucleosynthesis of \ce{^{14}N}, \ce{^{27}Al}, and, interestingly, \ce{^{39}K} within an accretion disk surrounding sub-solar mass black holes using the slim disk model. These possibilities arise when there is a reasonable accretion rate, close to the Eddington value, and an extremely low viscosity parameter. In the following section, we delve into a more detailed discussion regarding the impact of enrichment duration on the ejection of enriched material and examine the feasibility of such values for $m$ and $\alpha$.

Given the small duration of the enrichment in \ce{^{27}Al}, less than a few hundred seconds, even if a tiny black hole with mass $M = 0.01 M_{\odot}$ and an accretion disk with an extremely low viscous parameter ($\alpha < 10^{-3}$) exists, the ejection of the newly formed \ce{^{27}Al} would have to occur in an incredibly fine-tuned way in order to pollute the surrounding medium. We find the same results after examining the duration of the enrichment for \ce{^{39}K}, even when setting a smaller value for the enrichment ($\xi = 2$, justified by the range of [K/Fe] abundances being much smaller than the range of [Al/Fe], as shown in Fig.~\ref{fig:epsilon_in_context}).  

The low value for the viscous parameter $\alpha$ is a very challenging constraint for the nucleosynthesis. While numerical simulations suggest a value close to $\alpha \approx 0.02$ \citep{2011Hawley}, observations are better represented with $\alpha \approx 0.1$ \citep{2007King}. In any case, the values required to enrich the gas in \ce{^{14}N}, \ce{^{27}Al}, and \ce{^{39}K} is almost two orders of magnitude smaller than what is suggested by numerical simulations.

The last parameter putting strong constraints on the nucleosynthesis is the black hole mass, which also tends to be very small, although the mass required for the nucleosynthesis of \ce{^{14}N} is considerably more plausible compared to the cases of \ce{^{27}Al} and \ce{^{39}K}. When considering a modeling approach based on X-ray observations, a black hole with a mass of $1.3 M_{\odot}$ still falls within the range of very low likelihood in the distribution function of black hole masses in the Milky Way, as indicated by \cite{2010Ozel}.
Given our results on the black hole mass required for the nucleosynthesis, primordial black holes (PBHs) would be of particular interest. They are hypothetical astronomical objects, which would have formed shortly after the Big Bang, and are thought to have originated from the extreme density fluctuations during the early universe. They can possess a very wide range of masses, as suggested by \cite{2022PBH} and \cite{2023Carr}. For instance, the thermal-history-induced mass spectrum of PBHs peaks at $M \approx 1~M_{\odot}$ and has a lower tail reaching $M=10^{-8}~M_{\odot}$. \cite{2023Musco} also found, using numerical simulations, that the PBH mass function during the quantum chromodynamics epoch would peak at $M\approx 1 M_{\odot}$, with the lower mass tail extending until $10^{-3}~M_{\odot}$. They are dark matter candidates \citep{2021Carr} and evidence for their existence has recently been claimed using LIGO/Virgo gravitational-wave data \citep{2022Franciolini}. If such primordial black holes exist, they could be promising for globular clusters as they are among the oldest structures in the Universe. However, there are still many uncertainties around these objects such as:\ how long they survive across cosmic time; whether they can have an accretion disk and, if yes, what would be a suitable model to describe them; and whether they could have a much lower viscosity parameter, $\alpha$, that is compatible with the value found in this work. Undoubtedly, the progress of research on PBHs will be significantly propelled by upcoming gravitational wave telescopes such as Einstein and LISA \citep{2023Franciolini}.

\section{Conclusion} \label{sec:ccl}  
We have explored the possibility of accretion disks around stellar mass black holes as the polluter creating the enriched material at the origin of the phenomenon of multiple stellar population in globular clusters, following up on the study by \cite{2018Breen}.
Using the Python package pynucastro \citep{2022Clark}, we performed nuclear reaction simulations to investigate the condition in temperature and timescale required to create some of the anti-correlations between chemical elements observed in GCs. 

We find that a very small range of temperature and time (equivalent to density) could lead to the simultaneous \ce{^{12}C} - \ce{^{14}N}, \ce{^{16}O} - \ce{^{23}Na}, and \ce{^{24}Mg} - \ce{^{27}Al} anti-correlation patterns. However, the \ce{^{24}Mg} - \ce{^{39}K} anti-correlations observed in some GCs requires a different region of the parameter space. These results are in very good agreement with previous works \citep{2007Prantzos,2016Iliadis} and suggest that the enriched material is either coming from multiple polluters at different temperatures or from a single polluter having a large temperature range, for example black hole accretion disks.

In this study, we used the slim disk model with the framework outlined in the work by \citet{2006Watarai}. This optically and geometrically thick model proves to be particularly suitable for scenarios involving super-Eddington mass accretion rates. Here, advection emerges as a significant cooling mechanism and can no longer be neglected. Unlike the S-S model \citep{1973Shakura}, the slim disk model introduces less restrictive assumptions, notably by not imposing a Keplerian gas rotation profile. However, it is essential to acknowledge a potential limitation in our approach. To derive the volume density of the gas, an input for the nuclear reaction network, we employed a straightforward technique of dividing surface density by scale height. This method is typically employed under the one-zone approximation, which remains valid for geometrically thin accretion disks where the vertical variations of primary variables are averaged due to a significantly smaller scale height compared to the radial distance ($H << r$). However, this assumption does not hold true for the slim disk model. Nevertheless, given that the density is intricately linked with time in the nuclear reaction network and because the enrichment timescale generally exceeds the viscous timescale by several orders of magnitude, we anticipate that our results remain robust despite this assumption. Another limitation of our work lies in the absence of time dependence in the black hole accretion disk system. While using a constant value for the mass and viscosity is acceptable, the mass accretion rate can vary with time (see, e.g., \citealt{2004astro.ph.10551Vanderklis}). Depending on the timescale of the variation, this may or may not impact nucleosynthesis in the accretion disk.
Conducting further hydrodynamical simulations in conjunction with a nuclear reaction network within accretion disks could offer additional insights and constraints regarding the potential for nucleosynthesis. 

Using the slim disk model, we probed a very wide range of parameters for the black hole and accretion disk system, varying the black hole mass, $m$, the mass accretion rate, $\dot{m}$, the viscosity parameter, $\alpha$, and the radius, $r,$ over many orders of magnitude. Each combination of ($m$, $\dot{m}$, $\alpha$, $r$) leads to a temperature, $T$, density, $\rho$, and radial velocity, $v_R$, of the accretion flow. We then use $T$ and $\rho$ as input to the nuclear reaction network and obtain mass ratios for 132 chemical elements. We are particularly interested in chemical elements in over-abundance in the enriched stars of GCs, namely \ce{^{14}N}, \ce{^{23}Na}, \ce{^{27}Al}, and \ce{^{39}K}. 

We defined the enrichment timescale as the time needed for a chemical element to be enriched with respect to the pristine value, using a free parameter $\xi$ quantifying the level of enrichment. Subsequently, we compared the enrichment timescale to the viscous timescale, characterizing the lifetime of the gas before accretion onto the black hole, depending on the radius and the radial velocity of the flow. We generated 281,000 sets of values for the parameters ($m$, $\dot{m}$, $\alpha$, $r$)\ and we extracted 25,730 ($T$, $\rho$) pairs where the temperature is high enough for the nucleosynthesis to happen ($T>2\times 10^6$~K). Among theses pairs, only 82 lead to an enrichment timescale for \ce{^{27}Al} smaller than the viscous timescale, using an enrichment threshold of $\xi = 2$, meaning that \ce{^{27}Al} can be twice more enriched than the pristine mass ratio before the gas is accreted by the black hole in only $\approx 0.3\%$ of the simulated points. The number of points reduces to 20 for \ce{^{14}N}, to 6 for \ce{^{39}K}, and to 0 for \ce{^{23}Na}. The few quadruplets making an enrichment in \ce{^{27}Al} or \ce{^{39}K} possible all correspond to very light black holes $m<0.01 M_\odot$ and very low viscosity parameter $\alpha<10^{-3}$. Even under the assumption that stars consist entirely of enriched material ($\omega = 1$), it is important to note that such a low enrichment threshold ($\xi = 2$) will still be unable to recover extreme [Al/Fe] values.
The enrichment in \ce{^{14}N} requires a larger black hole mass, around $1~M_{\odot}$, but (as with \ce{^{27}Al} or \ce{^{39}K}) a very low viscosity parameter. Such a low value ($\alpha < 10^{-3}$) is disfavored by current observations and hydrodynamical simulations.
The question of whether  this kind of black hole could  really have existed in the past is yet to be answered.

\begin{acknowledgements}
We thank the anonymous referee for helpful suggestions that
improved the quality of our paper.
We thank Philip G. Breen for valuable suggestions and for reading the manuscript. We thank Prashin Jethwa for his careful reading of the manuscript and for providing useful comments.
Glenn van de Ven acknowledges funding from the European Research Council (ERC) under the European Union's Horizon 2020 research and innovation program under grant agreement No 724857 (Consolidator Grant ArcheoDyn). Elena Pancino acknowledges support from the INAF MS grant "Chemo-dynamics of globular clusters: the Gaia revolution (1.05.01.86.22)".
This research has made use of NASA’s Astrophysics Data System (ADS).
\end{acknowledgements}

%
\bibliographystyle{aa} 
\bibliography{BH_accretion_disk_FREOUR} 
%

\end{document}